\documentclass[aps,12pt,final]{revtex4b3}
\usepackage{graphicx}
\usepackage{dcolumn}
\usepackage{amsmath}

\begin{document}
\bibliographystyle{apsrev}

\title{Primary cosmic ray chemical composition in the
energy region around $10^{16}$~eV investigated by means of
$\gamma$-hadron families.}

\author{Maia Kalmakhelidze, Nina Roinishvili, Manana Svanidze}
\email{ninarich@hepi.iph.edu.ge}
\affiliation{E.Andronikashvili
Institute of Physics, Academy of Sciences of the Georgian
Republic, Tamarashvili 6, Tbilisi, Georgian Republic,\ 380077.}
\date{\today}
%\pacs{Cosmic ray, emulsion chamber}

\begin{abstract}
Primary Cosmic Ray Chemical Composition is investigated in energy\
region close to $10^{16}$ eV.  Studies are based on comparisons of
$\gamma$-hadron families observed by Pamir and Pamir-Chacaltaya
Collaboration, with families generated by means of quasi-scaling
model MC0. It is shown, that all characteristics of observed
families, including their intensity, are in a very good agreement
with simulated event properties at the normal chemical composition
and are in disagreement at heavy dominant compositions. Code
CORSICA with VENUS and DPM models also contradicts with
experimental data of families. One- and multi-dimensional methods
of recognition of Fe-like families is worked up and approved. They
are based on family characteristics  sensitive to atomic number of
induced nuclei and are not correlated between each others. It is
shown that the fraction of Fe-like families is consistent with the
normal chemical composition and strongly contradicts to heavy
dominant ones. The success of MC0 model, in description of
families properties, is due to large inelasticity coefficient of
soft interactions at superhigh energies.
\end{abstract}

\maketitle

\begin{center}
\chapter {\bf 1.~Introduction}
\end{center}
 The investigation of Primary Cosmic Ray (PCR) Chemical
Composition(CC) is one of the key problem for understanding of PCR
origin, properties of radiation sources and interstellar or extra
Galaxy space, which PCR passes from a source to the Earth. CC is
well investigated up to energy $10^{12}$eV. By direct methods CC
is studied up to energy ~$10^{15}$ eV \cite{1,2,3}. However the
reliability of results of these work is not great, since they base
on small statistics. At larger energies sources of the information
about CC is Extensive Air Showers (EAS) or families of
$\gamma$-quanta and hadrons, registered by X-ray emulsion chambers
. Despite of long-term researches the results are very
inconsistent. The statements concerning  CC swing from so-called
normal \cite{4,5} and even with proton dominant \cite{6} to heavy
\cite{6,7} and superheavy \cite{8} compositions.  The appropriate
data with the references are brought in Table ~\ref{tab1}.

Whereas the knowledge of CC in the region $10^{15}$ - $10^{16}$ eV
is especially important as a complete spectrum of PCR in this area
has a "knee".

The reason could be either spectrum bend of PCR (due to decrement
some the component from PCR content) or change  properties of
inelastic interaction between cosmic ray particles  and Earth's
atmosphere nucleus at these energies \cite{1,9}.

The present work is devoted to research of PCR CC in the energy
region directly after bend of a energy spectrum of PCR. The
  $\gamma$-hadron families are registered and
processed by Pamir and Pamir-Chacaltaya Collaborations. The
characteristic of the families are used for this task.

Running ahead we shall notice, that efficiency of families
generated by a nucleus, $\epsilon_A$, strongly depends on their
atomic number decreasing with increase of A. Therefore the
chemical composition of families differs much from CC of PCR:
\begin{eqnarray}
       f_A= \epsilon_A\times{C_A} / \Sigma ( \epsilon_A\times{C_A} )
%\label{1}
\end{eqnarray}
 For  comparison see Table~\ref{tab1} and \ref{tab2}.
\begin{table}
\caption{Used chemical composition of PCR at
 $E_0=10^{15}$~eV~, $C_A \%$}.\\
\label{tab1}
%\hspace{-4cm}
\begin{tabular}{|c|c|c|c|c|c|c|}
\hline~~ Composition~~&~~~~P~~~~&~~~~He~~~&~~CNO~~&~~SiMg
~~&~~~~Fe~~~~& $\langle \ln A\rangle $\\ \hline Normal\cite{4,5} &
40 & 20 & 10 & 10 & 20 &1.7\\ Heavy \cite{6,7} &15 & 10 & 17 & 0 &
58 & 3.0 \\ Superheavy \cite{8} & 7 & 5 & 12 & 6 &70 & 3.4 \\
 \hline
\end{tabular}
\end{table}

\begin{table}
\caption{Chemical composition of families, $f_{A}\%$}.\\
\label{tab2}
%\hspace{-6cm}
\begin{tabular}{|c|c|c|c|c|c|}
\hline ~~ Composition~~
&~~~~P~~~~&~~~~He~~~&~~CNO~~&~~SiMg~~&~~~~Fe~~~\\ \hline Normal
\cite{4,5}   & 75. & 16.6 & 3.2 & 2.4 & 2.8  \\ Heavy \cite{6, 7}
& 56. & 16.6 & 11. & 0 & 16.4  \\ Superveavy\cite{8}  & 40. & 12.6
& 12. & 4.4 & 31.\\ \hline
\end{tabular}
\end{table}

 Table~\ref{tab2} explains the main difficulty of investigation of
PCR CC based on data about families. Even if characteristics of
families induced by  heavy nuclei strongly differ from that of
proton families, influence of the former to the average
characteristics of families is not great, since CC of families is
strongly enriched by protons. Even at heavy composition of PCR, in
which iron makes about 60\%, they provide only 16.4\% of families.
How Table~\ref{tab2} is received and what is possible to make for the
analysis of PCR CC by data on families will be discussed in the
subsequent sections.

A $\gamma$-hadron family is a result of Nuclear Electromagnetic
Cascade (NEC)  developed in the Atmosphere after interaction of
PCR particles with nuclei of air in the top part of the
Atmosphere. This process is very complex, multistep and branching.
Therefore to get information from characteristics of families one
needs a comparison with Monte-Carlo calculation based on some mode
including both nuclear and electromagnetic cascades.

 Not less than ten of such models \cite{5,10,11,12,13,14,15,16,17}
  were developed up to day. All recent models of nuclear interactions
  \cite{5,15} are
quasi-scaling type. Scaling is violated in pionisation region. But
this almost not influences properties of families, since they are
formed mostly by particles from fragmentation region. Scaling is
violated a little in fragmentation region as long as the
interaction of a primary particle of PCR occurs not with a nucleus
of air. All these models are in the agreement with accelerator's
experiments and represent extrapolation of their properties to
superhigh energies. They differ insignificantly in details of the
interaction: in quantity of use flavors of secondary particles, in
a way of inclusion of diffraction processes and production of jets
with large transverse momentum.

In the majority of quasi-scaling models the intensity of families
at normal chemical composition is 2-4 times more then observed
\cite{18,19,20,21}. A dilemma is arisen: either admit scaling
violation in the fragmentation region or assume heavy composition
of PCR~\cite{7,8}. This explains the appearance of simulations
based on compositions with dominance of heavy elements
(Table~\ref{tab1}). By this point of view paper\cite{21} is
especially interesting as different possible decisions of the
dilemma are proposed in it.

MC0 model \cite{5} is used in the present work for the analysis of
data observed with the help of X-ray emulsion chambers. The Model
is based on the theory of quark-glion strings. Diffraction
processes, generation of jets with large transverse momentum,
production of strange and charm particles are included in it. The
important peculiarity of MC0 is a large inelasticity coefficient.
Details of the model are described in \cite{5}.

Let us note, that for to day the most popular is code CORSIKA with
some variants of strong interaction model. Comparisons of MC0
predictions  with results of CORSIKA calculations in variants DPM
and VENUS were made in work \cite{22,23,24}. There was shown that
MC0 predicts faster absorption of hadron component than both
variants of CORSIKA. As it will be apparent from the following
fast absorption of hadron component provides the agreement of MC0
model with observed intensity of $\gamma$-families. Model MC0 and
CORSIKA differs also in the average characteristics of families.
The characteristics in MC0 are closer to the experimental data
than in CORSIKA. This explains a choice of MC0 as base for the
subsequent analysis.

With the help of MC0 model nuclear-electromagnetic cascades in the
Atmosphere generated by various nuclei were simulated at power
like energy spectrum. Index of integral energy spectra of PCR,
$\gamma$, for all nuclei was also simulated. $\gamma$=1.7 for
energies less then $3\times10^{15}$~eV and $\gamma$=2.2 below the
"knee" were input in this case.

There are 6 sections in the present paper. Besides Introduction
(sec.1) and conclusion (sec.6) it contains 4 main parts each of
which are subdivided on two subsections concerning experimental
and modeling considerations. Section 2 is dealing with the problem
concerning intensity of families at the altitude of Pamir.
Observed (subsec.2.1) and predicted (subsec.2.2) intensities are
considered there. In the next two subsections (3.1 and 3.2)
properties of experimental and simulated families are described.
In subsection 4.1 the consistency of the model with the
experimental data is discussed while in subsection 4.2 this
subject is analyzed with the account of possible systematic errors
in the experiment. One- and multi- dimensional methods of
recognition of families generated by iron are elaborated and
applied to the Pamir data. This is done in subsections 5.1 and
5.2. 5.1 is a short review of the situation but 5.2 is an original
investigation. In the last section the results are summarized.

\begin{center}
\chapter{\bf2.~Intensity of  $\gamma$-hadron families }.
\end{center}

 In the next two subsections intensity of $\gamma$-hadron
families are studied from experimental and theoretical point of
view. In the end of subsection 2.2 the results are compared and a
solution of the dilemma, heavy composition or scaling violation in
the fragmentation region, is found. In the frame of quasi-scaling
model good agreement between observed and predicted intensity at
the normal chemical composition is achieved. This is due to large
inelasticity coefficient in MCO model.

\begin{center}
\chapter{\bf2.1~Intensity of  $\gamma$-hadron families, Pamir experiment.}
\end{center}

A group of $\gamma$-quanta (true $\gamma$ and also $e ^{+}$,
$e^{-}$) and hadrons ($\pi^{+}$, $\pi^{-}$, n, p...) produced in
NEC by interaction of a PCR particle with a nucleus of air is
called the $\gamma$-hadron family. In this paper we analyze
families satisfying the following conditions:
\begin{eqnarray}
100TeV\leq\Sigma E_{\gamma}\leq1000TeV,\nonumber \\ n_\gamma \geq
10,~~~ E_{\gamma},E_h^{\gamma} \geq 4TeV,~~~ 1cm<R_\gamma<15cm
%\label{2}
\end{eqnarray}

In (2) $E^{\gamma}_{h}$, $E_\gamma$ is visible energy of a hadron
or a $\gamma$-quantum, $R_\gamma$ - average radius of the family
count off from its energy weighed center.

Only $\gamma$-quanta seated at distance less than 15cm from the
energy-weighted centre of a family are included in it.

So-called, aggregation  procedure was used at processing
 of families. It consists of taking a pair of
 $\gamma$-quanta ($\gamma_{i} \gamma_{j}$), being at
 distance $R_{ij}$ smaller
 then 0.15 mm, as one with
 $E_\gamma=E_{\gamma_i}+E_{\gamma_j}$. The majority of
 such quanta actually combine their spots of darkness and
 the scanner's eye apprehends them as one. The rest we
 unite by the aggregation procedure as in real so in
 simulated families. Thus all $\gamma$-quanta with $R_{ij}<$0.15 mm
 are united.

Let us pay attention to the condition  $R_{\gamma}<$1cm. This
selection condition is applied to families for the first time. It
is connected with two circumstances. On the one hand the
characteristics of families with  $R_{\gamma}<$1cm are strongly
distorted by process of formation of dark spots on X-ray film of
emulsion chamber. A large part of them are overlapped, some of
them are almost completely united. These peculiarities of the
registration are difficult to reproduce in simulated families.
Therefore we prefer to exclude them from consideration. The same
reasons concern to families with $\Sigma{E_\gamma}\geq$ 1000 TeV.
In their central part there are very narrow bunches of
$\gamma$-quanta which are also difficult to separate. On the other
hand it was noticed \cite{25} that the relative part of narrow
families in experiment is more than in simulations (Fig. 1a) and
their properties differ from properties of the other families. For
example, families with $R_\gamma<$1cm in experiment are almost
away of hadrons and their spectrum of $E_\gamma$ is very similar
to the spectrum of $E_\gamma$ in purely electromagnetic cascades.
It also indicates that group of families with $R_\gamma<$1cm is
preferable to exclude from the analysis. Figure 1b shows that
after exclusion of events with $R_\gamma<$1cm experimental and
simulated distributions of $R_\gamma$  become closer.

Intensity of $\gamma$-hardon families~\cite{10} at Pamir level
(4370m above sea, 596 $gr/cm^{2}$) without condition
$R_{\gamma}\geq$1cm  is equal to
\begin{eqnarray}
(0.69 \pm 0.15)m^{-2} year^{ -1}str^{-1}
%\label{3}
\end{eqnarray}

Error bars include statistical and possible systematically
uncertainties.

 Total number of families studied in the
present work including narrow events ($R_\gamma<$1 cm) equals to
226. Without them it becomes 174.

Thus the experimental intensity of families satisfying the
criterion (2) is
\begin{eqnarray}
 I_{exp}=(0.69 \pm 0.15)\times174/226 =(0.53 \pm0.12)m^{-2}  year^{-1}str^{-1}
%\label{4}
\end{eqnarray}

\begin{center}
\chapter{\bf2.2~Intensity of $\gamma$-hadron families , MC0-model.}
\end{center}

The vertical intensity of~$\gamma$-hadron families can be
expressed as:
\begin{eqnarray}
         I_{fam}^{v}=\Sigma [I_A(E \geq E_0)\times\epsilon_A(E \geq E_0)\times\Omega_0 /\Omega_A^{fam}]
%\label{5}
\end{eqnarray}

 $I_{A}(E{\geq}E_0)$ - is an intensity of nuclei with atomic
number A and energy $E\geq{E_0}$,

$\epsilon _A(E \geq E_0)$ - is an efficiency of families
production satisfying  the given conditions.

In turn
\begin{eqnarray}
\epsilon _A = N^{A}_{fam} / N_A(E \geq E_0)
%\label{6}
\end{eqnarray}
 $N_{A}(E{\geq}E_0)$ is the number of NEC, induced by a
primary particle with atomic number A  and $N_{A}^{fam}$ - number
of families, which they have produced.

It is convenient to have $I_A$ in percentage of I, i.e. to
normalize it on total intensity:
\begin{eqnarray}
  C_A = I_A / I
%\label{7}
\end{eqnarray}
In (5) $\Omega _0 $ and $\Omega_{fam}$ are solid angles for
primary particles and families:
\begin{eqnarray}
\Omega _0=2\pi~~~~~\Omega_{fam}=2\pi/(1+T/\lambda_{att})
%\label{8}
\end{eqnarray}
 Here $T$ is air pressure at the
altitude of installation's exposition (600 gr/cm$^2$ for Pamir)
and $\lambda_{att}$ is attenuation mean free path of families.

As one can see from (5-8) at determination of calculated intensity
model defines only $\epsilon_A$ (efficiency of production),
whereas $C_A$ and the index of energy spectra of nuclei are set by
assumed chemical composition. Simulation calculations have shown
that $\lambda_{att}$ is independent of A in the limit of their
errors and equals to $75\pm 6 ~gr~cm^{-2}$. This value is in a
good agreement with the experimental figure $\lambda_{att}^{
exp}=78\pm 4 ~gr cm^{-2}$~\cite{10}.

Vertical intensity for all particles of PCR
\begin{eqnarray}
I^{v}(E \geq E_0) = \Sigma I_{A}^{v}(E \geq E_0)
%\label{9}
\end{eqnarray}
is estimated  up to very high energy. It is customary to use
empirical expression for  $\Sigma I_{A}^v(E \geq E_0) $ given in
article~\cite{4}. We also use it but change the integral energy
spectra index -1.6 accepted in ~\cite{4} to more recent value
$\gamma$=-1.7. As a result we have:

$I^v(E>10^{15} eV)=(50\pm 20)~m^{-2} year^{-1} srt^{-1}$

The new figure  for differential vertical intensity is obtained in
~\cite{1}. It gives for $I^v(E>10^{15})$:

$I^v(E>10^{15} eV) =(47\pm 12) ~m^{-2} year^{-1} srt^{-1} $.

Combining these two results we  use the following
value for $I^v$:
\begin{eqnarray}
  I(E>10^{15} eV)=(50\pm 15)~m^{-2}year^{-1} srt^{-1}
%\label{10}
\end{eqnarray}

Efficiencies  obtained by means of MC0 model for different nuclei
are brought in Table~\ref{tab3}.

$N_A^{sim}(E \geq E_0)$ are given in the first row of
Table\ref{tab3}. $N_A^{sim}(E \geq E_0)$ are actual number of
cascades
 simulated isotropically over zenithal angle interval from
$0^o$ up to $43^o$. For light nuclei ($P, He$) $E_0 = 10^{15}$ eV.
In the case of groups of nuclei $CNO$ and  $SiMg$ and also for
$Fe$ nuclei the minimal primary energy was put as
\begin{eqnarray}
E_0^A = A^{0.5} \times 10^{15}eV
%\label{11}
\end{eqnarray}
Such a choice of $\Theta_{max}$ and $E_0$ is determined by the
fact that nuclei with energy less than specified $E_0$ and $\Theta
> \Theta_{max}$ form no more than $2\%$ from total number of
families.

In the second row of Table\ref{tab3}  corrected number of cascades
$N_A(E \geq 10^{15}$ eV) are brought.

\begin{equation}
N_A(E \geq 10^{15} eV)=N_A^{sim}/(1-cos 43^o) / (E_0^A)^{1.7}
%\label{12}
\end{equation}

In Table~\ref{tab3} $N_A^{fam}$ is number of families generated by
the given nucleus and $\epsilon _A$ - efficiency of their
production. Data for protons having above the "knee" ($E_{knee} =
3\times10^{15}$ eV) spectral index $\gamma$ = -2.2 are given in
the last column of Table~\ref{tab3}.

Figures given in Tables~\ref{tab1} and~\ref{tab3} allow to
determine efficiency of family production, $\epsilon$, at the
given chemical composition, $C_A$, and energy spectrum index,
$\gamma$. Further on one can get predicted vertical intensity of
families, $I_{fam}^v$ , average energy of primary nuclei
responsible for families, $E_{fam}$, and average $\ln{A}$.
\begin{table}
\caption{Efficiency of family production by different nuclei,
$\epsilon_A$, and average energy of nuclei, responsible for them,
$E_A$.}
 \label{tab3}
 %\hspace{-1cm}
\begin{tabular}{|c|c|c|c|c|c|c|}
\hline
 &~~~~~P~~~~~&~~~~He~~~&~~~CNO~~~~&~~~~SiMg~~~~&~~~~Fe~~~&~~ P($\gamma$=-2.2)\\
\hline $N_A^{sim}\times{10^3}$  & 63.6 & 87 & 30 & 20 &  30 & 62
\\ $N_A \times{10^3}$  & 237 & 324 & 912 & 1113 & 3450 & 231 \\
$N_A^{fam}$ & 682 & 416 & 461 & 434 & 756 & 409\\ $\epsilon
_A\times100$  & $.29 \pm .01$ & $.13 \pm .01$ & $.051 \pm .003$ &
$.038 \pm .003$ & $.022 \pm .001$  & $.18 \pm .01$\\ $E_A$,PeV  &
15. & 29. & 53. & 66. & 80. & 9.\\ \hline
\end{tabular}
\end{table}
\begin{equation}
\epsilon = \Sigma (C_A \times \epsilon _A)/ \Sigma C_A
%\label{13}
\end{equation}
\begin{equation}
I_{fam}^v = I_0^v \Sigma [C_A \times \epsilon _A \times
(1+T/\lambda_{att}^A)]
%\label{14}
\end{equation}
\begin{equation}
E_{fam} = \Sigma (C_A \times \epsilon _A\times E_A) / \Sigma (C_A
\times\epsilon _A)
%\label{15 }
\end{equation}
\begin{equation}
\langle\ln{A}\rangle = \Sigma (C_A\times\ln{A}) / \Sigma C_A
%\label{16}
\end{equation}
The values $C_A$ and $\langle\ln{A}\rangle$ for investigated CC
are given in Table~\ref{tab1} (section 1), $\epsilon _A$ and $E_A$
- in Table~\ref{tab3}. Parameters calculated by  (13, 14, 15) are
brought in Table~\ref{tab4}. Dependencies of family production
efficiency on $\ln{A}$ for pure nuclei and on
$\langle\ln{A}\rangle$ for various chemical compositions are shown
in Figure 2a.  Predicted intensities of families at different
chemical compositions are brought in Figure 2b. In Figure 2b the
strip limited by broken lines corresponds to region allowed by the
experiment.  Table~\ref{tab4} and Figure 2b show that only normal
compositions (NC) with $\gamma$ = -1.7 for all nuclei and NC with
$\gamma$ = - 1.7 for all components except protons ($\gamma _P$ =
- 2.2 for energy more then $E_0 = 3\times10^{15}$~ eV) are
consistent with the experimental intensity of $\gamma$-hadron
families.  The heavy compositions ~\cite{7,8} predict intensity of
families essentially less then the experimental value.
 Let us note, that an account of possible systematic
errors in energy measurement of $\gamma$-quanta and hadrons only
strengthen reliability of the conclusions. Discussions in
subsection 4.2 show that predicted intensity at normal composition
become $(0.58 \pm 0.18) ~m^{-2} year^{-1} srt^{-1} $  and at heavy
composition - $(0.25 \pm 0.08) ~m^{-2} year^{-1} srt^{-1} $ after
the account for systematically errors.

Thus we come to  the following conclusions:
\begin{itemize}
\item
Model MC0 eliminates the dilemma: heavy chemical composition or
strong scaling violation in fragmentation region. In framework of
quasi-scaling models the agreement between experimental and
calculated intensity of families at normal chemical composition is
managed. This progress is due to rather large inelasticity
coefficient in MC0 model.
\item
  Heavy and superheavy chemical compositions in MC0 model  give too low
values of family's intensity.
\end{itemize}
One more result of simulations should be underline.
Table~\ref{tab3} indicates that the average energy of nuclei
     responsible to family's production is about $10^{16}$ eV.
Therefore the conclusion following from our investigations of
families correspond to the energy interval just above the "knee"
of energy spectra of PCR.
\begin{table}
\caption{Calculated parameters (see the text)}.\\ \label{tab4}
%\hspace{-3cm}
\begin{tabular}{|c|c|c|c|}
\hline
 ~~Composition ~&~ $~\epsilon\times100~$~ &~~~~~~ $I_{fam}$~~~~~~&~~$E_{fam}$,PeV~ \\
\hline Normal\cite{4} & 0.16 &$ 0.71\pm .22 $& 22.
\\Heavy\cite{6,7} & 0.076 &$ 0.35\pm .11 $& 32.  \\
Superheavy\cite{8} & 0.051 &$ 0.23\pm .07 $& 45.
\\Normal$(\gamma_p =-2.2)$  & 0.11 & $0.49\pm .15 $& 17.\\[1mm]
\hline
\end{tabular}
\end{table}

\begin{center}
\chapter{\bf3.~Characteristics of $\gamma$-hadron families.}
\end{center}

$\gamma$-hadron families are characterized by a number of measured
parameters. Their definitions and descriptions are given in the
next two subsections. Again in the first one experimental
questions and problems connected with measurement of parameters
are discussed. The results of calculations and some specific
circumstances are analyzed in the second subsection. On the base
of simulated events parameters sensitive to atomic number of
incident nucleus are found out. Short conclusions about agreement
between experimental and calculated values of parameters at the
normal composition are given in the end of subsection 3.2.

\begin{center}
\chapter{\bf3.1~ Characteristics of $\gamma $-hadron families, Pamir experiment.}
\end{center}

$\gamma$-hadron families are characterized by number of measuring
parameters. Conditionally they can be divided into 4 classes:
\begin{itemize}
\item
1.~Characteristics of $\gamma $-quanta related to energy:
$n_{\gamma }, \Sigma E_{\gamma } , \Sigma E_{\gamma }  / n_{\gamma
} $.

$n_{\gamma } $ - number of $\gamma  $-quanta, $\Sigma E_{\gamma }$
- total energy of $\gamma  $-quanta.
\item
2.~Spatial characteristics of $\gamma  $-quanta: $R_{\gamma },
E_{\gamma }R_{\gamma }, R_{\gamma }^E=\Sigma E_{\gamma }R_{\gamma
} / E_{\gamma }$ and $d$.

$R_{\gamma }$  - radius of a family (average distance from the
center of the family), $E_{\gamma }R_{\gamma }$ - average product
$E_{\gamma }R_{\gamma }$ , $R_{\gamma }^E$ average radius weighted
by energy.  Parameter $d = n_{in} / n_{obs}$ is defined as the
ratio of the number of initial $\gamma $-quanta, $n_{in}$, to the
number of observed $\gamma $-quanta, $n_{obs} = n_{\gamma}$.  An
initial $\gamma $-quantum is responsible for a narrow group of
spots on an X-ray film, which are the result of an electromagnetic
cascade induced by it in the Atmosphere.  Observed dark spots
being on a small distance, $R_{ij}$, from each other are combined
into one initial $\gamma $-quantum, if $R_{ij} / (1/E_i + 1/E_j) <
10~TeV\times{mm}$. Using this algorithm, called decascading
procedure, the number of initial $\gamma $-quanta is determined.

\item
3.~Characteristics of hadrons related to energy:  $n_h$, $\Sigma
E_h^{\gamma}$,

 $q_E=\Sigma E_h^{\gamma}/(\Sigma E^{\gamma}, + \Sigma E_h^{\gamma})$,
$q_n=n_{\gamma}/(n_{\gamma}+n_h) $

$n_h$ - number of hadrons, $E_h^{\gamma }$ - energy transferred by
a hadron into soft component (visible energy of a hadron), $\Sigma
E_h^{\gamma}$ - total visible energy of hadrons, $q_E $ - fraction
of energy carried by hadron component of a family and $q_n$ -
fraction of hadrons in total multiplicity.

\item
4.~Spatial characteristics of hadrons: $R_h, E_h^{\gamma} R_h$ and
etc. Let us note, that the last characteristics are not examined
in this work, since the number of hadrons is, as a rule, small and
consequently their spatial characteristics have very wide
fluctuations.
%\end{enumerate}
\end{itemize}
\begin{table}
\caption{Average values of parameters, P, their statistical
errors, $\sigma_P$, in experimental families and sensitivity of
parameters, S.} \label{tab5}
%\hspace{-3cm}
\begin{tabular}{|c|c|c|c|c|c|c|c|}
\hline
~~~~~~~~~~&~~$n_h$~~&~~~$R_\gamma$(cm)~~~&~~~$R_\gamma
^E$~~ &~~ $~E_\gamma R_\gamma ~$ &~~~$d
$~~~&~~~$q_n$~~~&~~$q_E$~~~\\
%& & & & & &\\
\hline P &~3.1~ &~2.8~ & 2.4~&~27.~ &~.63~ &~.11~~&~~.14~ \\
$\sigma _P$ &~0.3~ &~0.1~ & 0.1~&~2.0~ &~.01~ &~.01~~&~~.01~\\
\hline S &~1.38~ &~1.33~ & 1.26~&~1.04~ &~.96~ &~.65~~&~~.60~ \\
\hline
\end{tabular}
\end{table}

Parameters belonging to a given class are subject to common
systematic errors. In the 1-st class they are determined by errors
in energy measurement of $\gamma $-quanta.  For energy from 4 up
to 50 TeV the relative error is about 20$\%$.  The effects of
saturation of darkness and the overlapping of spots appeared for
larger energies. Whenever possible these effects are taking into
account during primary processes of families. At the next stage of
treatment, $\gamma $-quanta being on distance $R_{ij} < 0.15$ mm
are united in one and only families with $\Sigma E_{\gamma} <
1000$ TeV and $R_{\gamma} > 1$cm are included into analysis (see
subsection 2.1).  As it seems these restrictions eliminate the
main part of systematic errors of the 1-st class parameters.  This
question is discussed in more detail in subsection 4.2.
Complexities in determination of spatial characteristics (the 2-nd
class) were already discussed in subsection2.1. We should remind
that aggregation of quanta with $R_{ij} < 0.15$ mm and exclusion
of families with $R_{\gamma} < 1$ cm from analysis set aside most
difficulties.  Main uncertainties of the 3-rd class's parameters
are connected with determination of visible energy of a hadron,
$E_h^{\gamma}$, based on its darkness.  Due to two reason the
situation here is easier than with $\gamma$-quanta: the saturation
of a darkness comes later, as distribution of density is more flat
in case of hadron and overlapping is not present, since average
distance between hadrons is much more than between
$\gamma$-quanta.  However, the relation between visible energy and
density of darkness  for hadron are investigated not so carefully
as for $\gamma$-quanta~\cite{26}. This complexity will be
discussed in subsection3.2 in connection with determination of
visible energy of hadrons in simulated families.  Average values
of parameters of the experimental families with their statistical
errors are given in Table~\ref{tab5}.

Sensitivity of a parameter to atomic number of primary
 particle is defined as:
\begin{equation}
            S=(<P_{Fe}>-<P_P>)/D_P
%\label{17}
\end{equation}

$<P_{Fe}>$ - is an average value of the given parameter for
families induced by $Fe$, $<P_P>$ - the same for families induced
by protons, $D_P$ - dissipation of a parameter for primary proton.
$S$ was calculated by means of simulated families for primary
protons and iron.  Let us note, that all parameters are defined in
such a way that $<P_P>$ is less than $<P_{Fe}>$. For this purpose
in two cases it was necessary to depart from initial definitions
of parameters. Parameter $d$ was introduced for the first time in
work ~\cite{27} as the ratio $d = n_{obs} / n_{in}$. We have
redefined it by replacing $d \to 1/d$. Parameter $q_E$, was
introduced in work ~\cite{28} as $q_E=\Sigma E_{\gamma}/(\Sigma
E^{\gamma}, + \Sigma E_h^{\gamma})$. We transformed it as $q_E \to
1-q_E$.  In Table~\ref{tab5} parameters are brought in order of
decrease of their sensitivity. Characteristics of families with
$S<0.5$ are not given there. At a research of chemical composition
they can be only harmful. Not having sensitivity they are useless
but the systematic errors in them can enter distortions into final
results. Distributions of two parameters sensitive ($R_{\gamma} $)
and not sensitive ($n_{\gamma }$) to atomic number of a primary
nuclei are demonstrated in Figures 3a and 3b.

\begin{center}
\chapter{\bf3.2~ Characteristics of $\gamma $-hadron families, MC0-model.}
\end{center}

The average value of a given parameter of families, P,
 at certain
chemical composition can be expressed by a formula:
\begin{eqnarray}
P= \Sigma (C_A\times \epsilon_A\times P_A)/ \Sigma (C_A
\times\epsilon_A)
%\label{18}
\end{eqnarray}
where $P_A$ is an average value of the given parameter in families
generated by a nucleus with atomic number $A$; $C_A$ - a fraction
of nuclei $A$ at the given chemical composition of PCR; $\epsilon
_A$ - an efficiency of family production by nuclei $A$ with the
energy above $10^{15} $eV.  As above, the model determines values
$P_A$ and $\epsilon _A$, whereas CC and the power index of energy
spectra of nuclei $A$ are set as an input of simulations.  It was
already noted that all procedures used at selection and processing
of experimental data were applied to simulated families.
Peculiarity of simulated events is a modelling of hadron
registration in X-ray emulsion chamber and determination of energy
transferred by it into the soft component, $E^h_{\gamma} =
K_{\gamma}\times{E_h}$. A hadron is registered by X-ray film if it
has interacted in the chamber and energy transferred by it,
$E^h_{\gamma}$, is more than threshold value - 4~TeV. The
probability of hadron interaction and factor $K_{\gamma}$ is
determined by a design of X-ray emulsion chamber. For Pamir carbon
chambers special investigations ~\cite{29} have shown that the
probability of interaction is about 0.7 and $K_{\gamma}$ has a
distribution, $f(K_{\gamma})$, similar to incomplete
$\gamma$-function
\begin{eqnarray}
f(K_\gamma)=
A\times{K_\gamma^\alpha}\times{exp~(-K_\gamma/\beta)},
 ~~\langle{K_\gamma}\rangle=(\alpha+1)\times{\beta}
%\label{19}
\end{eqnarray}
For example, if $\alpha=1.5$ and $\beta=0.075$ average
$<K_{\gamma}>$ has quite reasonable values equal to 0.188.
However, the condition $E_h^{\gamma}>4$ TeV makes $<K_{\gamma}>$
dependent on $E_h$ (Figure 4a).

Therefore the more important parameter is not
$\langle{K_\gamma}\rangle$ but $K_{eff}$
\begin{eqnarray}
K_{eff}= \int \int K_\gamma f(K_\gamma) F(E_h) dE_h dK_\gamma
%\label{20}
\end{eqnarray}

Here $F(E_h)dE_h$ is the energy spectrum of hadrons in families.
As calculations shown for simulated families $K_{eff}=0.23$. This
value is quite compatible with that offered in ~\cite{29}.
Distributions of $K_{\gamma}$ obtained in ~\cite{29} for Pamir
chamber and our approximation of $f(K_{\gamma})$ are given in
Figure 4b. They are in a reasonable agreement.  Knowing
probability of hadron interaction, its energy and distribution of
$K_{\gamma}$ and using Monte-Carlo method one gets for each hadron
answers to questions: whether hadron interact in the chamber? what
energy it transfers into soft component
$E_h^{\gamma}=(K_{\gamma}E_h)$? and whether it is registered, i.e.
$E_h^{\gamma}$  is more than $E_{thr} = 4$ TeV? Such algorithm was
applied to simulated families to get hadron characteristics: $n_h,
q_n, q_E$, etc.  Average values of sensitive ($S>0.5$) parameters
of families, $P_A$, and their dispersions, $D_P$, for various
primary nuclei($P^{\alpha}$~-~Protons with energy spectrum having
"knee")
  are brought in Table~\ref{tab6}.

\begin{table}
\caption{Average values of parameters, $P_A$, and their
dissipation, $D_P$, in simulated families.} \label{tab6}
%\hspace{-4cm}
\begin{tabular}{|c|c|c|c|c|c|c|c|}
\hline
 ~~~ &~~~ $~n_h~$~~ &~~~ $R_\gamma $~~~(cm)~~&~~~$R_\gamma ^E$ ~~&~~ $~E_\gamma R_\gamma ~$
&~~~$d $~~~&~~~$q_n$~~~&~~~$q_E$~~~\\ \hline P &~2.7~ &~2.8~ &
2.4~&~24.~ &~.59~ &~.10~~&~~.10~ \\ $D_P$ &~2.4~ &~1.5~ &
1.6~&~15.~ &~.16~ &~.07~~&~~.11~ \\\hline $P^{~\alpha}$  &~2.5~
&~2.7~ & 2.3~&~24.~ &~.60~ &~.10~~&~~.10~
\\ $D_P $ &~2.4~ &~1.4~ & 1.5~&~14.~ &~.16~ &~.08~~&~~.10~ \\ \hline He
&~3.3~ &~3.2~ & 2.9~&~30.~ &~.63~ &~.12~~&~~.20~ \\ $D_{He}$
&~2.8~ &~1.6~ & 1.7~&~17.~ &~.16~ &~.08~~&~~.11~ \\\hline CNO
&~4.3~ &~4.0~ & 3.6~&~34.~ &~.69~ &~.14~~&~~.15~ \\ $D_{CNO}$
&~3.2~ &~1.7~ & 1.9~&~17.~ &~.13~ &~.08~~&~~.10~ \\\hline SiMg
&~5.0~ &~4.5~ & 4.1~&~38.~ &~.72~ &~.16~~&~~.16~ \\ $D_{SiMg}$
&~4.0~ &~1.8~ & 2.0~&~19.~ &~.13~ &~.08~~&~~.11~ \\\hline Fe
&~6.3~ &~4.8~ & 4.4~&~40.~ &~.74~ &~.18~~&~~.19~ \\ $D_{Fe}$
&~4.7~ &~1.9~ & 2.1~&~18.~ &~.13~ &~.09~~&~~.11~ \\ \hline
\end{tabular}
\end{table}
Dependencies of $n_h , R_{\gamma}$ and $d$ on $A$ are shown in
Figures 5a and 5b.  As above characteristics of families for all
nuclei are calculated at integral energy spectrum index $\gamma
=-1.7$. Data for families generated by protons having power
spectrum with bend in point $E_0=3\times{10^{15}}$ eV are brought
in the third and fourth lines. Up to the bend $\gamma =-1.7$,
after $\gamma =-2.2$.  Table\ref{tab6} shows that the "knee" of
the spectrum insignificant influences on average characteristics
of proton induced families.
\begin{table}
\caption{Expected values of families characteristics at various
chemical compositions.} \label{tab7}
%\hspace{-2cm}
\begin{tabular}{|c|c|c|c|c|c|c|c|}
\hline
~~~~~~~~~ &~~ $~n_h~$~~ &~~~ $R_\gamma $~~~&~~~$R_\gamma
^E$ ~~~~&~~ $~R_\gamma E_\gamma ~$ &~ ~~$d
$~~~&~~$q_n$~~&~~$q_E$~~\\ \hline Normal &~3.2~ &~3.0~ &
2.6~&~26.~ &~.62~ &~.11~~&~~.11~
\\Heavy &~3.6~ &~3.3~ & 2.9~&~29.~ &~.64~ &~.12~~&~~.12~ \\
Superheavy &~4.2 ~ &~3.7~ & 3.3~&~32.~ &~.66~ &~.13~~&~~.14~ \\
\hline
\end{tabular}
\end{table}

 One can calculate expected values of parameters
for the given CC (Table~\ref{tab7}) with the help of expression
(18) and using Tables~\ref{tab1},~\ref{tab3} and~\ref{tab6} where
various compositions, $C_A$, efficiency of family's generation,
$\epsilon _A$, and the average values of parameters, $P_A$, are
given.

Comparison of Tables~\ref{tab5} and~\ref{tab7}  shows that
characteristics of families at normal composition are in good
agreement with the experimental data whereas predictions for heavy
and the more so for superheavy compositions differ much from the
observations. This subject is discussed in the next two
subsections.

\begin{center}
\chapter{\bf4.~Comparison of the experimental data with the results of MC0
model.}
\end{center}

Detailed comparison of the experimental data with the results of
MC0 model is given in the following two subsections. In the first
one $\chi ^2$ test is applied to the above mentioned sensitive
parameters but before parameters uncorrelated between each others
were found out. All of them show a much better agreement with the
results corresponding to the normal chemical composition than that
at the heavy compositions. In the next subsection not only
sensitive but other main parameters of families are attracted to
the analyze with the aim of study the possible systematic errors
in the experiment. Here also good agreement with calculations at
the normal chemical composition is achieved.

\begin{center}
\chapter{\bf4.1~Compatibility of the experimental data with the results of
MC0 model.}
\end{center}

Before making the final conclusions about chemical composition of
PCR by comparison of experimental data (Table ~\ref{tab5}) with
predictions of MC0 model (Table~\ref{tab7}) one has to find out
which of the studied characteristics are not correlated. Otherwise
traditional (for example, $\chi ^2$) and not traditional (neural
net) approaches can incorrectly estimate a degree of agreement of
experimental and simulated data.  Among selected seven parameters
sensitive to chemical composition (Table ~\ref{tab6}) there are 2
groups of strongly correlated characteristics. They are
characteristics related to the energy of hadrons, $n_h, q_n, q_E $
(for example see Figure 6a) and spatial characteristics of
$\gamma$-quanta, $R_{\gamma}, R_{\gamma}^E, E_{\gamma}R_{\gamma} $
(Figure 6b). Stands to apart parameter $d$, which weekly correlate
with both of the groups (Figure 6c).  Parameters belonging to the
different classes do not correlate also (Figure 6d). We have
chosen three sensitive and not correlated parameters: $n_h$,
$R_{\gamma}$, and $d$.  Only they are used in the subsequent
analysis.  For comparisons of the experimental and simulated
families the following quantities were calculated:
\begin{equation}
\chi ^2_p= [(P_{exp} -P_{mod}) / \sigma P_{exp}]^2
%\label{21}
\end{equation}
and sum for three not correlated parameters
\begin{eqnarray}
\chi ^2_3= {\bf [} [(n_{h exp} - n_{h mod}) / \sigma n_{h exp}]^2
+ ~~~~~~~~~~~~~~~~~~ \nonumber \\ ~~~[(R_{\gamma exp}- R_{\gamma
mod})/ \sigma R_{\gamma exp}]^2+ [(d_{exp}- d_{mod})/ \sigma
d_{exp}]^2 {\bf ]}~ /~3~~~~~~~~~~~~~~~~~~~
%\label{22}
\end{eqnarray}
Here $P_{exp}$ and $P_{mod}$ are an average value of some
parameter but $\sigma P_{exp}$ is an error of $P_{exp}$.  Instead
of total error $(\sigma _{exp}^2+\sigma _{mod}^2)$ only
corresponding $\sigma _{exp}^2$ stands in  (21) and (22) since
statistical errors in calculations are much less than experimental
uncertainties. The results are shown in Table~\ref{tab8}.

Each $\chi ^2$ should be near one if the experiment and
calculations are in a good agreement, since the number of degrees
of freedom for separate parameter is 1 and $\chi ^2_3$  is an
appropriate $\chi ^2$ with the 3 degrees of freedom divided on 3.
This expectation is fulfilled only for the normal composition, but
is not satisfied for the heavy and superheavy compositions. It is
necessary to notice, that rather large values of $\chi
^2_{R\gamma}$ =3.4 for the normal composition can indicate on the
presence of some systematic errors. This problem will be analyzed
in the next subsection.  The dependence of $\chi ^2_3$  on $<lnA>$
is shown in Figure 7. Dotted line corresponds to $\chi ^2_3$ =4.
Confidence level at $\chi ^2$ =4 is less than 1$\%$. Figure 7
shows that all compositions with $ <lnA> $ more than 2.5 are above
the line and therefore can be excluded with confidence level more
than  99$\%$.
\begin{table}
\caption{Values $\chi ^2_p$  for various parameters and $\chi
^2_3$.} \label{tab8}
%\hspace{-6cm}
\begin{tabular}{|c|c|c|c|c|}
\hline ~~~~~CC ~~~~& ~~~$\chi ^2_{nh}$~~ &~~~ $\chi ^2_{R\gamma}
$~~~&~~~$\chi ^2_d$~~~ &~~$\chi ^2_3$ \\ \hline Normal
&~0.11~&~3.4~ & 0.95~&~1.5~  \\ Heavy &~2.7~ &~20.~ & 0.24~&~7.7~
\\ Superheavy &~13. ~ &~54.~ & 7.2~&~25.~
\\ \hline
\end{tabular}
\end{table}

\begin{center}
\chapter{\bf4.2~Consents of MC0 model with the experimental data.
Systematic errors.}
\end{center}

It was shown in the previous section that MC0 model with normal
chemical composition of PCR is in a reasonable agreement with
experimental data. Nevertheless two circumstances cause doubt.
Firstly, experimental average
    radius of families $R_{\gamma} = 2.8 \pm 0.1$ appears to be less than
calculated radiuses at any chemical compositions $R_{\gamma}$ =
3.0 - 3.7 (Table~\ref{tab5} and~\ref{tab7}). This is the cause of
a little bit large values $\chi ^2_{R\gamma}$=3.4 and accordingly
$\chi ^2_3$=1.5 for the normal CC (Table~\ref{tab8}). For heavy
compositions these quantities are much more.  Secondly,
experimental average values of $n_{\gamma} = 21 \pm 2$ is less
than calculated $n_{\gamma} = 24 - 26$ at any chemical
composition. Till now $n_{\gamma}$ was not interesting as it does
not depend on atomic number of incident nuclei (see Figure 3b).
There is a suspicion that $R_{\gamma}$ and $n_{\gamma}$  are
subject of some systematic experimental errors and it is desirable
to find their common origin.  In this subsection we want to carry
out the more complete analysis of the consent of MC0 model with
experimental data considering possible systematic errors at
measurement of parameters. Values of the most important parameters
characterizing $\gamma$-hadron families measured in the our
experiment and predicted by MC0 model for various CC together with
$\chi ^2_P$ are brought in Table~\ref{tab9}.
\begin{table}
\caption{Values of parameters, $P$, and their $\chi ^2_P$.}
\label{tab9}
%\hspace{-3cm}
\begin{tabular}{|c|c|c|c|c|c|c|c|c|}
\hline
Exp and CC ~~&~~ Param.~~ & ~~$~n_h~$~~ &~~~ $R_\gamma
$~~~&~~~~$d$~~~&~~~~$q_n$~~~~& $~E_\gamma R_\gamma ~$
&~~~$n_\gamma $~~~~&~~~~$E_ \gamma$ \\ \hline Exp& ~~~P & 3.1& 2.8
& .63 & .11 & 27. & 21. & 12.1 \\
          & $ \sigma _P$& 0.3 & .01 & .01 & .01 & 2. & 2. & 0.5 \\\hline
Normal & ~~~P & 3.2& 3.0 & .62 & .11 & 26. & 24. & 11.1 \\
          & $ \chi ^2_P$& 0.11& 3.4 & .95 & .06 & 1.3& 13. & 3.1 \\\hline
Heavy    & ~~~P & 3.6& 3.3 & .64 & .12 & 29. & 25. & 10.8 \\
          & $ \chi ^2_P$& 2.71& 20. & .24 & 1.5 & 0.86& 20. & 7.0 \\\hline
Superheavy& ~~~P & 4.2& 3.7 & .66 & .13 & 32. & 26. & 10.5 \\
          & $ \chi ^2_P$& 13.& 54. & 7.2 & 9.5 & 5.2& 25. & 12.4\\
\hline
\end{tabular}
\end{table}
As already was noted the most serious deviation of model from
experimental data is displayed in parameters $R_{\gamma}$  and
$n_{\gamma}$. For heavy composition this disagreement is even
stronger than at the normal CC. The common reason of all large
deviations can be systematic errors in energy determination,
$E_{\gamma}$, by measurement of spot's darkness, $D$, on a X-ray
film. The algorithm of transition from $D$ to $E_{\gamma}$ is
rather complex. Besides theoretical connection between energy of
$\gamma$-quantum and the number of electrons in a circle of
certain radius under given thickness of lead where a film is
located (transition from $E_{\gamma}$ to $n_e$) it includes the
property of the film ($n_e \to D$ curve, saturation of $D$ at
large $n_e$ etc), condition of the exposition of the film (its
background, gap between the film and lead plate) and so on. All
these effects mostly are accounted and controlled.  However, it is
not excluded that some of discrepancy can nevertheless be left.
Group effects in families are added to these problems - partial
overlapping of spots and even their aggregation (the last effect
we took into account at processing of simulated families). To
study an influence of energy measurement errors on family's
characteristics we have added to simulation of events modeling of
distortion by replacing $E_{true}$ to $E_{meas}$.

For this we used some distortion functions, $f(E)$:
\begin{equation}
f(E_{true}) = E_{meas}/{E_{true}}
%\label{23}
\end{equation}
Two distortion functions were tested :

\begin{equation}
f_1(E_{true}) = 10/E_{true}\times(E_{true}/10)^\alpha
%\label{24}
\end{equation}
with $\alpha >$1 ($\alpha<$1) up to 10~TeV and $\alpha <$1
($\alpha>$1)
 after 10~TeV to provide underestimation (overestimation) of true
energy

and
\begin{equation}
f_2(E_{true}) = 70/E_{true}\times(E_{true}/70)^\beta
%\label{25}
\end{equation}
 with $\beta<$1 in the whole interval of $E_{true}$. Parameters  $\alpha$
and $\beta$ were varied.  Typical examples of distortion functions
are shown in Figure 8. At the particular chose of $\alpha$,
function $f_1(E)$ reflects underestimation of energy up to and
after $E_{true}$=10~TeV with different dependence of $E_{meas}$ on
$E_{true}$. Opposite to $f_1(E), f_2(E)$ overestimates true energy
up to 70~TeV and then underestimates it. $f_2(E)$ was taken from
work\cite{30}. An attempt of modeling of passage of
$\gamma$-quanta through the lead plates of the chamber and then
its fixing on a film with the accounts of peculiarities mentioned
above was carried out there.  Various values of $\alpha$ were set
in calculations to ensure $10 \%, 20 \%, 50 \%$, (-$20 \%$ and
-$40 \%$) underestimation (overestimation) of energy near measured
threshold 4~TeV and $-5 \%, - 10 \%, -20 \%$ underestimation of
energy at true energy 100TeV. Dependencies of $\chi^2_P$ for the
normal chemical composition on a degree of energy underestimation
or overestimation in Figure ~9a ,9b.  The Figures show that all
$\chi ^2_P$ have quite admissible values already at $10\%$ of
underestimation of energy ($E_{true}$ = 4.4~TeV instead
$E_{meas}$= 4~TeV) and no systematic errors at 100TeV ($E_{meas} =
E_{true} $ at $E_{true}
>$100~TeV). It doesn't means absence of systematic errors at large
energy $E_{true} >$100~TeV.  Simply it appears that values $\chi
^2_P$ weekly depend on an underestimation of energy at $E_{true}
>$100~TeV (Figure 9b).  As the $10\%$ systematic error is quite
possible we conclude once more that MC0 model at the normal
chemical composition is in a very good agreement with the
experimental data.  Situation at heavy and the more so at
superheavy compositions is different.  In the previous section it
was shown that they contradict to the experiment. Figure 9c,
similar to 9a, for heavy composition shows that no one distortion
function can improve the consent of MC0 model with the experiment
at heavy composition. $\chi ^2_P$ for normal and heavy chemical
compositions at the most favorable distortion functions for each
of them are given in Table~\ref{tab10}. We should note that in the
case of superheavy composition $\chi ^2_P$ values are so large
that this composition is out of consideration.

\begin{table}
\caption{$\chi^2_P$ values for normal and heavy chemical
compositions at 10$\%$ and 20$\%$ underestimation of energy near
the threshold.} \label{tab10}
%\hspace{-1cm}
\begin{tabular}{|c|c|c|c|c|c|c|c|c|c|}
\hline
~~~~ CC ~~&~~ $E_{true}$~~ &~~ $~n_h~$~~ &~~~ $R_\gamma
$~~~&~~~~$d$~~~&~~~~$q_n$~~~~&~~ $~E_\gamma R_\gamma ~$
~~&~~~$n_\gamma $~~~~&~~~~$E_ \gamma$~~~&~~~ $\chi ^2_3$ \\ \hline
Normal & ~4.4 & 0.09& 1.85& .11 & .02 & 0.11& 3.4 & 1.37 & 0.70\\
\hline Heavy    & ~4.8 & 0.63 & 7.39&2.65 &1.84 & 1.88& 4.63& 2.15
& 3.60\\
 \hline
\end{tabular}
\end{table}

Finally let us discuss the predicted intensity of families at used
distortion functions. Since after introduction of distortion
functions the number of events, satisfying the selection criteria
of families, decreases, diminished also expected intensity of
families. For the normal composition at $E_{true}$ = 4.4~TeV the
intensity become equal to $I_{fam}= (0.58 \pm 0.18)$ m$^{-2}$
year$^{-1}$ srt$^{-1} $ .  For the heavy composition at $E_{true}$
= 4.8~TeV $I_{fam}= (0.25 \pm 0.08)$ m$^{-2}$ year$^{-1}$
srt$^{-1} $. Let us remind that experimental intensity is (0.43
$\pm$ 0.12) m$^{-2}$ year$^{-1}$ srt$^{-1} $.  From the given
figures it is apparent that after corrections for possible
systematic errors the conclusions made in section 2 concerning
chemical compositions based on intensity of families strengthen.

We have used a number of procedures simulating process of
$\gamma$-quanta and hadrons registration in X-ray chamber. These
are: a procedure of aggregation (unification near-by spots); an
ascription to each hadron a factor of energy transfer into soft
component, $K_{\gamma}$, distributed according to incomplete
$\gamma$-function; possible underestimation of energy threshold of
$\gamma$-quanta and hadrons, $E_{thr}$; an account for systematic
underestimation of energy of the most energetic $\gamma$-quanta
(replacement in simulations $E_{true}$ to $E_{meas}$). This can
appear to be not correct and artificial. Actually, as any complex
installation, X-ray emulsion chamber brings in certain distortions
in NEC falling on it. As a result instead of true parameters,
$P_{true}$, one has to do with their measured values, $P_{meas}$.
To have an opportunity to compare parameters measured in
experiment with that of simulated, it is necessary to model
processes occurring in X-ray emulsion chamber. In Pamir
collaboration the routine, carrying out this problem is named the
code of "passages through the chamber"\cite{30}. This code is
rather complex and is an analogue of routine JEANT \cite{31}, used
at accelerators for imitation of distortions, connected with this
or that installation: ATLAS, CMS and others\cite{32,33}. In the
present work "the program of passage through the chamber" was
divided on stapes (aggregation, $K_{\gamma}, E_{true} \to
E_{meas}$) to look after influence of each of the effects
separately. By such approach we have resulted in the conclusions
about a role of each type of distortions in determination of
measured parameters of families.

\begin{center}
\chapter{\bf5.~Selections of families generated by iron nuclei.}
\end{center}

 The attempt of selections of families similar to that induced
 iron(Fe)nuclei by means of image recognition methods were done in the
several papers. All of them claim that compositions enriched with
heavy components are in contradiction to experimental results.
Only compositions near to the normal can satisfy them. The
description of the method and short review of the mentioned works
are given in the  subsection 5.1. In the next subsection our
original investigations are brought. The most encouraging is the
result of reconstruction  the fraction of Fe-like families
inputted into calculations by the method, which is proposed. The
main conclusion is again – only results at the normal composition
are consistent with the experiment.

\begin{center}
\chapter{\bf5.1~Selections of families generated by iron nuclei.  A short
review.}
\end{center}

The attempt to determine a fraction of iron nuclei in PCR on a
database of X-ray-emulsion chamber was undertaken in work
\cite{27, 34, 35} for the first time. Almost simultaneously by two
groups of authors it was offered to apply to family
multidimensional analysis of an image recognition. The sense of it
is  that using a difference in distributions of some parameter of
families generated by iron and proton (see Fig. 3a) one defines a
limiting value for the given parameter, $P_{lim}$. Event with
$P_{fam} \geq {P_{lim}}$, is considered as a family similar to one
generated by iron. Such an event is attributed to group of
families "Fe". Otherwise if $P_{fam} < P_{lim}$, then family  is
added to group "P". A fraction of families, $f'$, " similar to
iron " is thus defined. Multidimensional method of an image
recognition means that the limiting values are defined for several
parameters $P_{lim}^{1}$, $P_{lim}^{2}$, $P_{lim}^{3}$ ....
Families which simultaneously satisfy the requirements:
\begin{center}
$P^{1} \geq {P_{lim}^{1}}$,~~~~ $P^{2} \geq {P_{lim}^{2}}$...
\end{center}

are attributed to group "Fe".

Fraction of families induced by  nuclei A  and satisfying the
limiting conditions ( Fe-like produced events) are designated
$R_{A}$.

Two types of errors in the image recognition method. The error of
the first type is to attribute families from proton to  "Fe" group
($R_P$ ) and the error of the second type is not to "recognise"
iron family  and not to  added it to  "Fe" group, ($1 - R_{Fe}$).
The quality of selection is defined by values of these two errors.
As more sensitive are the parameters to atomic number, the
stronger differ distributions of parameters for P and Fe families.
As a consequence cleaner and more complete is the selection as
both errors are small.  The error of the first type includes all
families from other primary nuclei (except Fe) falsely attributed
to the "Fe" group.

Quasi-scaling models were used for determination of limiting
values of parameters $P_{lim}~^{i}$ in  three mentioned works.
Training sets of families generated by P and Fe were simulated and
distributions, similar to that shown in Figure 3, were obtained.
These distributions allowed determination of the boundary
quantities $P_{lim}~^{i}$. It was found~\cite{27, 34} that it is
impossible to use simultaneously more than two parameters because
of limited statistics of families. Therefore either
$E_{\gamma}R_{\gamma}$ and asymmetry parameter of family, $b$, or
$E_{\gamma}R_{\gamma}$ and $n_{\gamma}$ were used in~\cite{34}.
Another parameter of asymmetry, $\alpha$, and parameter $d$, used
in the previous sections, or their combination with $R_{\gamma}$
were involved in ~\cite{27}. Parameters $d$, $\alpha$ and
$1/R_{\gamma}^{E}$ in different combinations were analyzed in
~\cite{35}.

The fraction of families attributed to group "Fe", $f'$, is equal
to
\begin{equation}
f ' = f \times R_{Fe} +(1-f) \times R_{P}
%\label{26}
\end{equation}
where $f$ is the real fraction of families generated by iron. From
(26)
\begin{equation}
f = (f ' - R_{P} ) / (R_{Fe} +R_{P})
%\label{27}
\end{equation}
Authors of the all three works~(\cite{27,34,35}) applying the
described method to families of Pamir Collaboration have found
that the fraction of families generated by iron does not exceed
2-3 $\%$, i.e. that the compositions of PCR with domination of
iron contradict to the experiment.

The method of selection families generated by iron was essentially
improved in work ~\cite{6}, where experimental data of
Pamir-Chacaltaya Collaborations were used. In contrast to us (see
conditions (2)) in papers based on Joint  Pamir-Chacaltaya
experiment a group of $\gamma$-quanta and hadrons with
$\Sigma{E_{\gamma}} +\Sigma{E_h^\gamma}
>$100~TeV at $E_{\gamma}$ and $E_h^\gamma >$ 4~TeV
was considered as a family ~\cite{6}. For recognition of an image
of a family produced by iron author of [6!!] used a neural net
method, with the help of which multi-dimensional analysis is
reduced to one-dimensional. 15 parameters describing a family were
used in ~\cite{6}: $n_{\gamma}$, $n_h$, $\Sigma{E_\gamma}$,
$\Sigma{E_h}$  and so on. With the help of neural net the set of
them was reduced to one, $y_p$, and condition which attributes a
family to the "Fe" group was  $y_p
>$0.5. The quality of selection
making by this way has appeared to be rather high: $R_P\approx {(1
- R_{Fe})}\approx{15 \%}$.

Also the next step on the way of CC study was made in ~\cite{6}:
not only families produced by P and Fe but also generated by other
nuclei (He and CNO, SiMg groups) were examined there. It means
that these nuclei also contribute to group "Fe". Following our
consideration in this case(24) should be transformed to:
\begin{equation}
f ' = \Sigma (f_A \times {R_A})
%\label{28}
\end{equation}
and accordingly
\begin{equation}
f = (f '- \Sigma' {(f_A \times{  R_A)}}) / R_{Fe}
%\label{29}
\end{equation}
Here $f_A$ is a fraction of families generated by nuclei A, $R_A$
- their fraction faulty attributed  to  group  "Fe"
(except~$R_{Fe}$, $R_{Fe}$ is true portion), $f '$ - the part of
families satisfying the limiting condition. In $\Sigma(f_{A}
\times{R_{A}})$ contributions of all nuclei are summarized, in
$\Sigma'(f_A \times{R_A})$ iron families are not included. Let us
underline once more that we designate true value of Fe induced
families as $f_{Fe}$ and estimated by (28) as $f$. The training
sets define $R_A$ and the given chemical composition of PCR
determines $f_A$ (see (1) and Table~\ref{tab2}.

The conclusion of \cite{6} is the same as in previous works
\cite{34, 27,35}: the heavy composition of PCR is excluded by
experimental data concerning $\gamma$-hadron families.

For the sake of completeness of the review refer to work~
\cite{8}. In this work rather complex parameters, reflecting
structural properties of families, and intensity of "structural"
families are used. The conclusion in ~ \cite{8} is opposite to the
above-stated. Authors affirm that the superheavy composition of
PCR is necessary for description of considered properties. As one
of the critical remarks to ~ \cite{8} let us note that authors of
~ \cite{8} did not show that the model they used describes the
more simple characteristics of families at the superheavy
composition.

\begin{center}
\chapter{\bf5.2~Selections of families generated by iron nuclei. Original
consideration.}
\end{center}

We have used methods similar to multi-dimensional \cite{27,34} and
one-dimensional \cite{6} approaches  as a following step of
chemical composition researches. Before starting CC analysis we
have found  parameters $n_{\gamma}$, $R_{\gamma}$, and $d$
sensitive to A and are not correlated between each other (section
3). In contrast to \cite{6, 27,34} we used only them. The
parameters are analyzed in their reduced form:

\begin{equation}
X_{P} = (P - P_{p}) / \delta{P_p}
%\label{30}
\end{equation}

Here $P$ is a value of some parameter in a given family either
experimental  or simulated, $P_P$ - average values of the same
parameter in families induced by  protons, $\delta{P_P}$ -
disipation of this parameter in proton families.

In reduced form of variables all distributions are dimensionless
and for families generated by protons are dispersed around average
values $X_P$ = 0. At multidimensional analysis simultaneously
three conditions were used to attribute a family to "Fe" group:
\begin{equation}
X_{n_h}> X_{n_h ~lim},~~~ X_{R_\gamma}>X_{R_\gamma~ lim},~~~ X_d>
X_{d~lim}
%\label{31}
\end{equation}
For one-dimensional analysis a new parameter $X_{3}$, was
introduced:
\begin{equation}
X_3=(X_{n_h}+  X_{R_\gamma}+ X_d)/3
%\label{32}
\end{equation}
For this parameter a value $X_{3~ lim}$  was also found and the
family was attributed to group "Fe" if its
\begin{equation}
X_3  > X_{3~ lim}
%\label{33}
\end{equation}
Limiting conditions $P_{lim}$  were determined with the help of
integral distributions of $X_{P}$ for families induced by P and Fe
using training sets of simulations. Integral distributions at once
show what fraction of proton families ($R_{P}$) at the given
$P_{lim}$ will be by mistake attributed to "Fe" group and what
fraction of iron families will not recognized as Fe-like,
(1-$R_{Fe}$). An example of such distribution is given in Figure
10. The arrows in Figure 10 indicate limiting values of $X_{3}$.
One of them corresponds to about equal errors $R_{P} = (1-R_{Fe})
\approx{20\%}$, while the other - to  $R_{P} = 5 \%$ and $(1 -
R_{Fe})\approx 40 \%$.

It is apparent that both errors $R_P$ and $1- R_{Fe}$ depend on
chosen limits $X_{n_h~lim}$, $X_{R_{\gamma}~lim}$, $X_{d~lim}$,
and $X_{3~lim}$. We investigated two sets of limiting parameters.
First, at which $R_P = 5\%$, i.e. only $5\%$ of proton families
are falsely attributed to "Fe" group, and second, such that the
errors of the first and second types are approximately equal $R_P
\approx (1-R_{Fe})$. True values of  fraction iron induced
families, $f_{Fe }$(see Table~\ref{tab2}), and expected fraction
of families, selected to group "Fe", $f '$, for different chemical
compositions are brought in Table~\ref{tab11} at two sets of
limiting parameters $X_{P ~lim}$. Let us remind that $f_{Fe }$ is
fully determined by the given CC while $f ^{'}$ is the result of
processing of simulated data for set of limiting parameters.

\begin{table}
\caption{True fraction of Fe families, $f_{Fe}$, and expected
fraction of Fe-like families,$f ^{'}$,at various chemical
compositions}.\\ \label{tab11}
%\hspace{-2cm}
\begin{tabular}{|c|c|c|c|c|}
\hline
  ~~~ &~~~~~ CC~~~~~ &~~ $f_{Fe}\%$,~true ~~~&~~~ $f^{'}\%$,~one-dim. ~~&~~ $f^{'}\%$,~multi-dim.
  \\
  \hline
   & Normal & 2.8 & 13. & 11. \\
  $R_P=5\%$ & Heavy & 16.4 & 21. &19. \\
   & Superheavy & 31. & 30. & 28. \\\hline
   & Exp. &  & 12.$\pm$3. & 11.$\pm$3. \\
  \hline\hline
   & Normal & 2.8 & 26. & 27. \\
  $R_P=20\%$ & Heavy & 16.4 & 35. & 37. \\
   & Superheavy & 31. & 47. & 47. \\\hline
   & Exp. &  & 29.$\pm$4. & 25.$\pm$4. \\
  \hline
\end{tabular}
\end{table}

Table~\ref{tab11} is composed using training sets for various
nuclei P, He, CNO, SiMg and Fe.  $R_A$ was determined for each
nucleus and then the sum  $f ^{'}= \Sigma{(f_{A} \times R_A)}$ was
found. We should like to pay attention to the fact that for normal
composition the true iron families figural spiking is  lost among
families selected in  "Fe" group. At the normal composition the
true Fe fraction  $f_{Fe}= 2.8\%$ while  Fe-like proton $f^{ '}$
varies from  11$\% $-to 27$\%$ depending on $R_{P}$.

To check efficiency of the used method to determine a part of Fe
induced families we have  investigated   how it reproduces
$f_{Fe}$ at various chemical compositions. The results ($f$) are
given in Table~\ref{tab12} for one particular selection
(multi-dimensional selection, $R_P=20\%$). Results of the other
three selections (multi-dimensional $R_P = 5\%$, one-dimensional
at $R_P=20\%$ and $5\%$) are identical.
\begin{table}
\caption{Fractions of iron induced families, $f$ , in chemical
composition of families, calculated by (29). Multi-dimensional
selection, $R_P = 20\%$}.\\ \label{tab12}
%\hspace{-2cm}
\begin{tabular}{|c|c|c|c|c|c|c|}
\hline
 ~~~~~&~~~~~P~~~&~~~~ Normal~~ & ~~Heavy~~ & ~~Superheavy~~ &~~ Fe ~~&~~ $f_{Fe}$,$\%$ \\\hline
  P & \bf{0.2$\pm$2.5} & -4.3$\pm$ & -2.6$\pm$ & -1.2$\pm$ & 27.$\pm$2.5 & 0 \\
  Normal & 8.7$\pm$1.7 &\bf{4.2$\pm$1.7}& 5.9$\pm$1.7 & 7.3$\pm$1.7 & 36.$\pm$1.7 &2.8 \\
  Heavy & 21.$\pm$2.4 & 16.$\pm$2.4 &\bf{18.$\pm$2.4} & 19.$\pm$2.4 & 48.$\pm$ & 16.4 \\
  Superheavy & 35.$\pm$2.7 & 30.$\pm$2.7 & 32.$\pm$2.7 & \bf{34.$\pm$2.7} & 62.$\pm$2.7 & 31. \\
  Fe & 79.$\pm$6.4 & 75.$\pm$6.4 &76.$\pm$6.4& 78.$\pm$6.4 & \bf{106.$\pm$6.4} & 100  \\
   &  &  &  &  &  &  \\\hline
  Exp. & 5.9$\pm$5.0 &\bf{1.4$\pm$5.0} & 3.1$\pm$5.0 & 4.5$\pm$5.0 & 33.$\pm$5. &\bf{ ?} \\
  \hline
\end{tabular}
\end{table}

In Table ~\ref{tab12} rows correspond to simulated compositions,
columns to compositions by  means of which corrections were done
using (29) . For comparison $f$ with the true values of $f_{Fe}$
the letter for each simulated compositions are given in the last
column of the Table.

Table ~\ref{tab12} demonstrates rather encouraging result.
Independently to the composition used for corrections (for
exception of pure Fe), the fractions $f$ are close to
corresponding true value.  As it is expected the best agreement
$f$ with true $f_{Fe}$ is obtained if the composition for
corrections is close to the "real". The corresponding figures are
underlined in Table ~\ref{tab12}. From the above the following
procedure of processing of experimental data is suggested. After
determining an experimental value $f ^{'}$ corrections should be
done for various CC, for example for five compositions testing in
Table~\ref{tab12}. Then receiving preliminary result (five values
for $f$) one takes that at which chemical composition determining
corrections is closest to obtained $f$. This is shown in the last
row of Table ~\ref{tab12}. The final value of found fraction of Fe
induced families is underlined.

The total analysis of our experimental data was as follows. Four
values of fractions of families similar to iron, $f^{ '}$, were
found corresponding to two type of the methods (multi-dimensional
and one-dimensional) and two sets of limiting parameters for
$R_{p}=5\%$ and $R_{p}=20\%$. They are brought in the last row of
Table~\ref{tab11}. Values $f$ were reproduced by the help (29).
Quantities $f$ are given in the last row of Table ~\ref{tab12}
corresponding to various correction compositions. Table
~\ref{tab12} shows, that only normal composition gives self
consistent values $f$. In this case true values $f_{Fe}=2.8\%$ and
set of experimental quantities of $f$ fluctuates from 1.4$\%$ to
4.5$\%$. In the case of the heavy composition the experimental
fraction $f=3.1\%$ contradicts to true value $f_{Fe} =16.4\%$.
Even more disagreement shows the superheavy composition: true
$f_{Fe} = 31\%$ while obtained f is equal to 4,5$\%$.

As it was specified above we have tested four selections criteria
of Fe-like families. Table~\ref{tab12} is given for one of them.
The results of all other selections are identical: events,
generated by iron nuclei, compose about (2 - 3)$\%$. This is in
agreement with  the normal composition and sharply contradicts to
compositions enriched by iron.

\begin{center}
\chapter{\bf6.~Conclusion}
\end{center}

The purpose of the present work was an investigation of
 Chemical Composition of Primary Cosmic Rays with energy
  around $10^{16}$~eV close to the "knee" of energy
  spectrum of PCR. In this region of energy information
  about CC of PCR can be obtained only by an indirect
  way on the base of study of extensive air showers or
  families of $\gamma$-quanta and hadrons registered by
  X-ray emulsion chambers.

In the present work the data of Pamir and Pamir-Chacaltaya
 Collaborations concerning families were used. The analysis
  of a material was made by means of comparison of the
  experimental families with that simulated by
  quasi-scaling model MC0~\cite{5}.

First of all it was shown that MC0 at normal CC (close to
 composition at $10^{14}$eV, about 40$\%$ P and 20$\%$ Fe)
 predicts intensity of families in
 complete agreement with experimental observations
 (section 2). Thus long-term dispute, heavy chemical
  composition or strong scaling violation in the
  fragmentation region was solved.

Further it was shown that not only intensity, but also all
 main characteristic of families (they are about 15) are
  well described by MC0 at normal chemical composition
  (section 3).

Characteristics of families sensitive to atomic number of
an incident nucleus were found out on the base of
 simulated events generated by primary protons and iron.
  Among them 3 parameters not correlating with each others
   were chosen. They are number of hadrons $n_{h}$, radius of
   a family $R_{\gamma}$ and parameter $d$, describing
   electromagnetic
   structure of an event. It has appeared that their
   average values in experiment are close to the same in
   artificial families at the normal composition of PCR and
     are in disagreement with the heavy (15$\%$ P and 58$\%$ Fe)
      and,
    especially, with  the superheavy (7$\%$ P and 70$\%$ Fe)
    compositions. $\chi_3^2$ was calculated for the three above
     mentioned parameters. In the case of the normal CC it is
      equal to 1.9, for two iron enriched CC $\chi_3^2$= 7.7 and
       25. correspondingly (subsection4.1).

Despite of the rather good agreement of experiment and
calculations at the normal composition two basic characteristics
of families $n_{\gamma}$ and $R_{\gamma}$ have rather significant
deviations from expected values: $\chi_{n_\gamma}^{2}$=13. and
$\chi_{R_\gamma}^{2}$= 3.4. The investigations have shown  that if
one admits 10$\%$ underestimation of energy near to registration
threshold of $\gamma$-quanta (about 4~TeV) and introduces the
appropriate distortion function into simulated families, then
corresponding $\chi^{2}$   become equal to $\chi_{n_\gamma}^{2}$=
3.4 and $\chi_{R_\gamma}^{2}$=1.85 (subsection4.2). At 20$\%$
underestimation of energy $\chi^{2}$  for all characteristics of
families are close to 1. On the other hand any underestimation or
overestimation of energy do not bring to an agreement between
experimental average characteristics of families to those
simulated at heavy and superheavy compositions.

Another attempt to investigate chemical composition was based on
the selection of families similar to that generated by iron. Such
method was already applied to experimental data of Pamir
\cite{27,34,35} and Pamir-Chacaltaya  Collaborations~\cite{6}.In
all four investigations fraction of Fe-generation $\gamma$-hadron
families was small and equal 2-3$\%$ . Thus it appear that the
heavy compositions contradict to the fraction of families from Fe
in the experimental data.

In this article we continued elaboration of an application of
image recognition methods to CC research by means of families
(sections 5). Limiting values ($X_{P~ lim}$) for the three above
mentioned sensitive parameters (P) in their reduced form $X_{P}$
were determined (section 5). Families with $X_{P} > X_{P~ lim}$
were attributed to "Fe" group. A united parameter $X_{3} =
(X_{n_\gamma} + X_{R_\gamma} +X_{d})/3$  with its $X_{3~lim}$ was
also used. Ability of the method was investigated by the help of
simulated families. Sets of families generated by pure protons and
irons as well as by the normal, heavy and superheavy chemical
compositions were analyzed. Families "similar" to iron were
selected for all these sets. The fractions of Fe-like families and
the restored ((29)) parts of "true" Fe families were estimated.
For each set of simulations the found "true" fraction of
Fe-families were in coincidence with the fraction of Fe families
in the given chemical composition.

The method of selection of Fe-like families was applied to
experimental events. A fraction of families similar to that
induced by iron was found. It appeared to be close to the
appropriate fraction at the normal composition. After corrections
by means of (29) the part of families from iron in the
experimental set constitutes  (2 - 3)$\%$. Such value is in
complete agreement with the normal CC and in sharp contradiction
to the iron enriched compositions.

Thus the basic conclusions of the given work are:

MC0 model at the normal chemical composition completely agrees
with the experimental data of $\gamma$-hadron families;

The chemical composition of Primary Cosmic Ray in the energy
region near to $10^{16}$~eV just above the "knee" of its energy
spectrum is close to the chemical composition at energy around
$10^{14}$~eV;

The models of nuclear electromagnetic cascade in the Atmosphere
with chemical compositions enriched by heavy elements contradict
to the experimental data on families. They predict too low
intensity of families, incorrect values of the characteristics of
families and too large fraction of families generated by iron.

One more conclusion very  important for the particle physics
follows from the present work. All the quasi-scaling
models~\cite{6,12,14,15,22,23,24}, except MC0, give 2-3 times
larger intensity of families than the observed one. It means that
in these models hadron absorption mean free part is too large to
provide proper intensity of families. Smaller mean free part in
MC0 is due to large an inelasticity coefficient  than in the other
models. The better agreement of average characteristics of
families in MC0 than in other models ~\cite{22,23,24} is also the
result of stronger absorption of hadrons. Therefore large an
inelasticity coefficient  is necessary feature of inelastic
interactions at superhigh energies.

\begin{acknowledgments}
The authors express their gratitude to all participants of Pamir
Collaboration for long and fruitful joint work, R.Mukhamedshin for
granting an opportunity to use the code MC0 developed by him and
also M.Tamada for making available the data bank of
Pamir-Chacaltaya Collaboration. Very useful were discussions with
Dr.'s R.Mukhamedshin and M.Tamada.
\end{acknowledgments}

%\newpage
\begin{figure}
\includegraphics{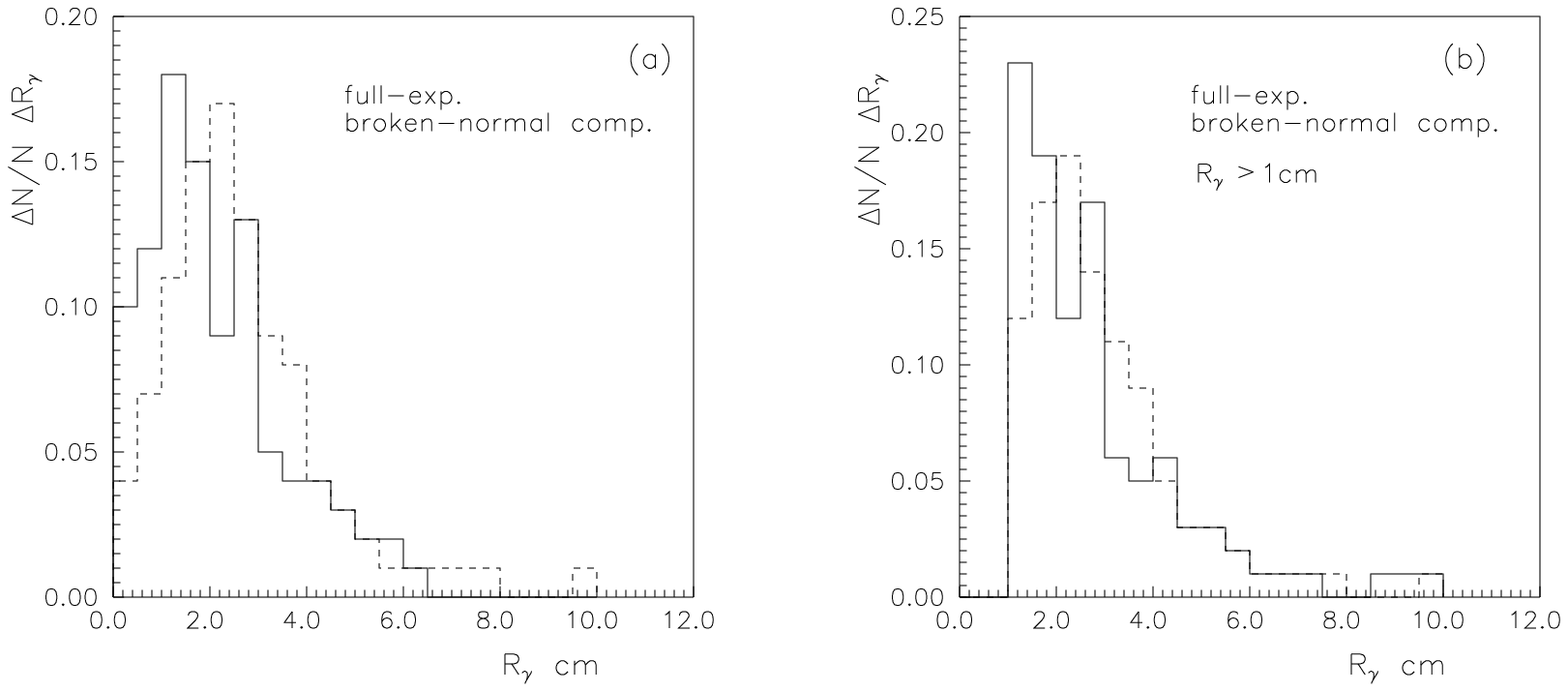}
\caption{$R_{\gamma}$ - distributions. Continues line -
experiment, dashed - normal CC. a) all families, b) families with
$R_{\gamma} > $1cm. }
\end{figure}
%!!!!!!!!!!!!!!!!!!!!!!!
\begin{figure}
\includegraphics{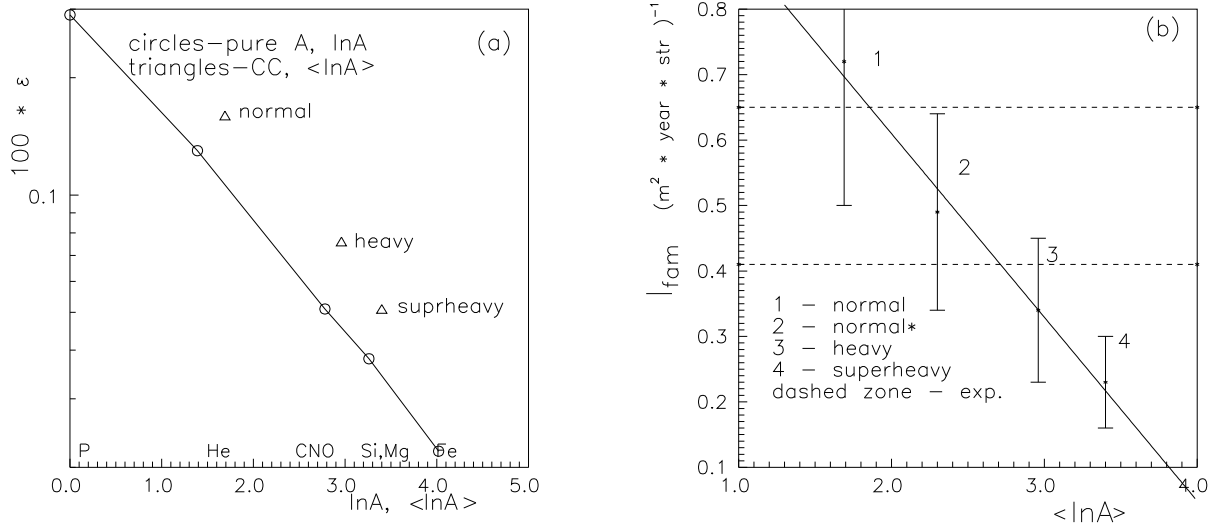}
\caption{a)  Families  production efficiency as function of
$\ln{A}$. o - pure nuclei, $\Delta$ - different CC. b)  Intensity
of $\gamma$-families  at different CC. }
\end{figure}
\begin{figure}
\includegraphics{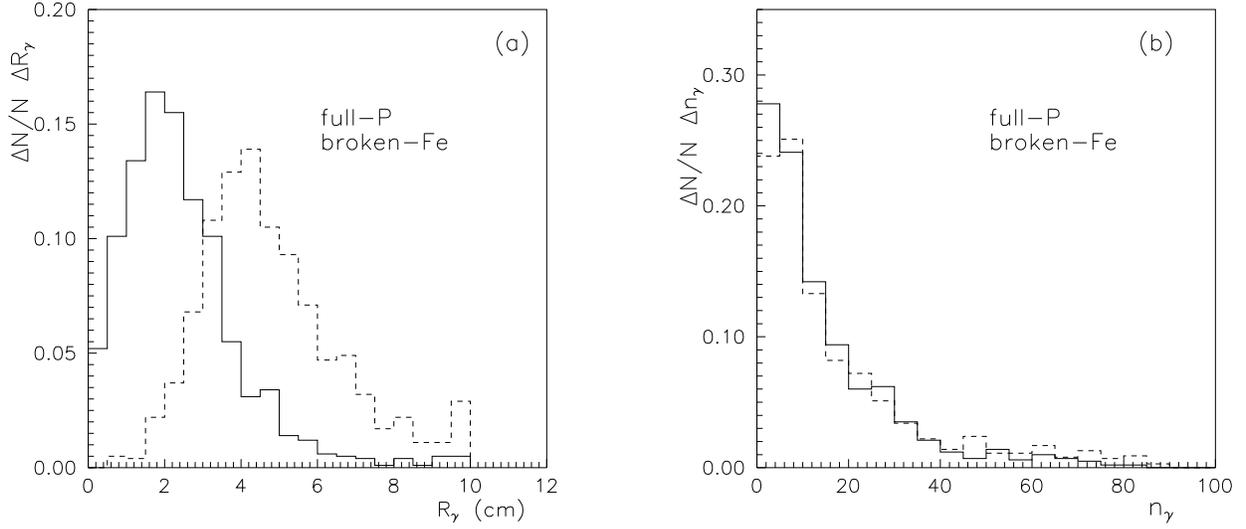}
\caption{Distributions of $R_{\gamma}$ (a) and $n_{\gamma}$ (b)
for families generated by protons (continues line) and iron
(dashed line). }
\end{figure}

\begin{figure}
\includegraphics{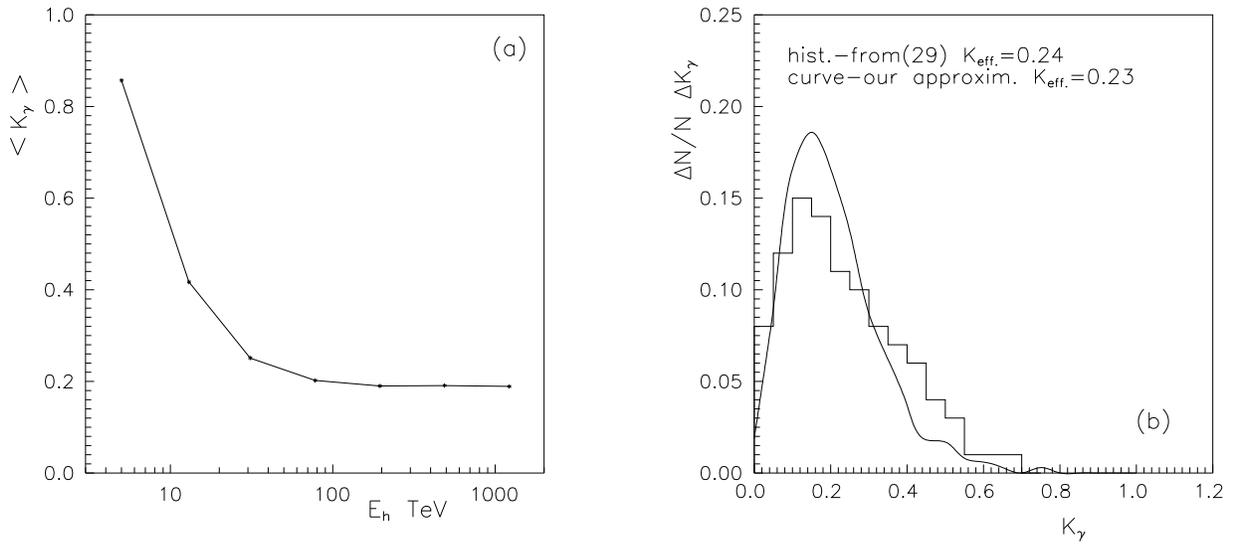}
\caption{a) Dependence of $\langle{K_\gamma}\rangle$ on $E_{h}$.
b) $K_{eff}$ distribution. Histogram is taken from ~\cite{29},
curve is used in this work. }
\end{figure}

\begin{figure}
\includegraphics{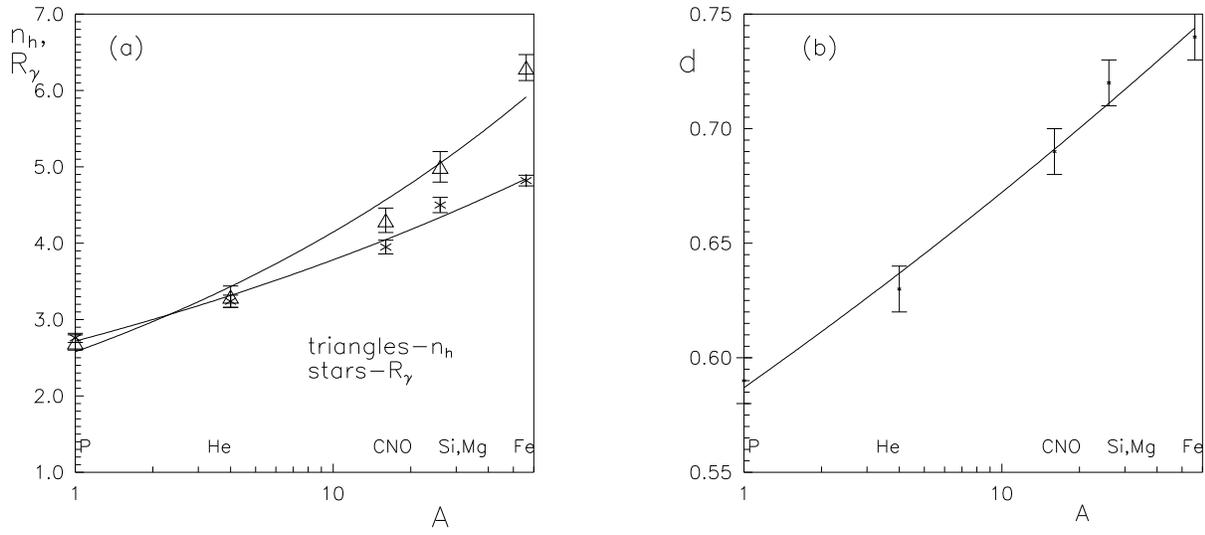}
\caption{a)$n_h$ and $R_{\gamma}$  dependencies on A. b)
dependence of $d$ on A.}
\end{figure}

\begin{figure}
\includegraphics{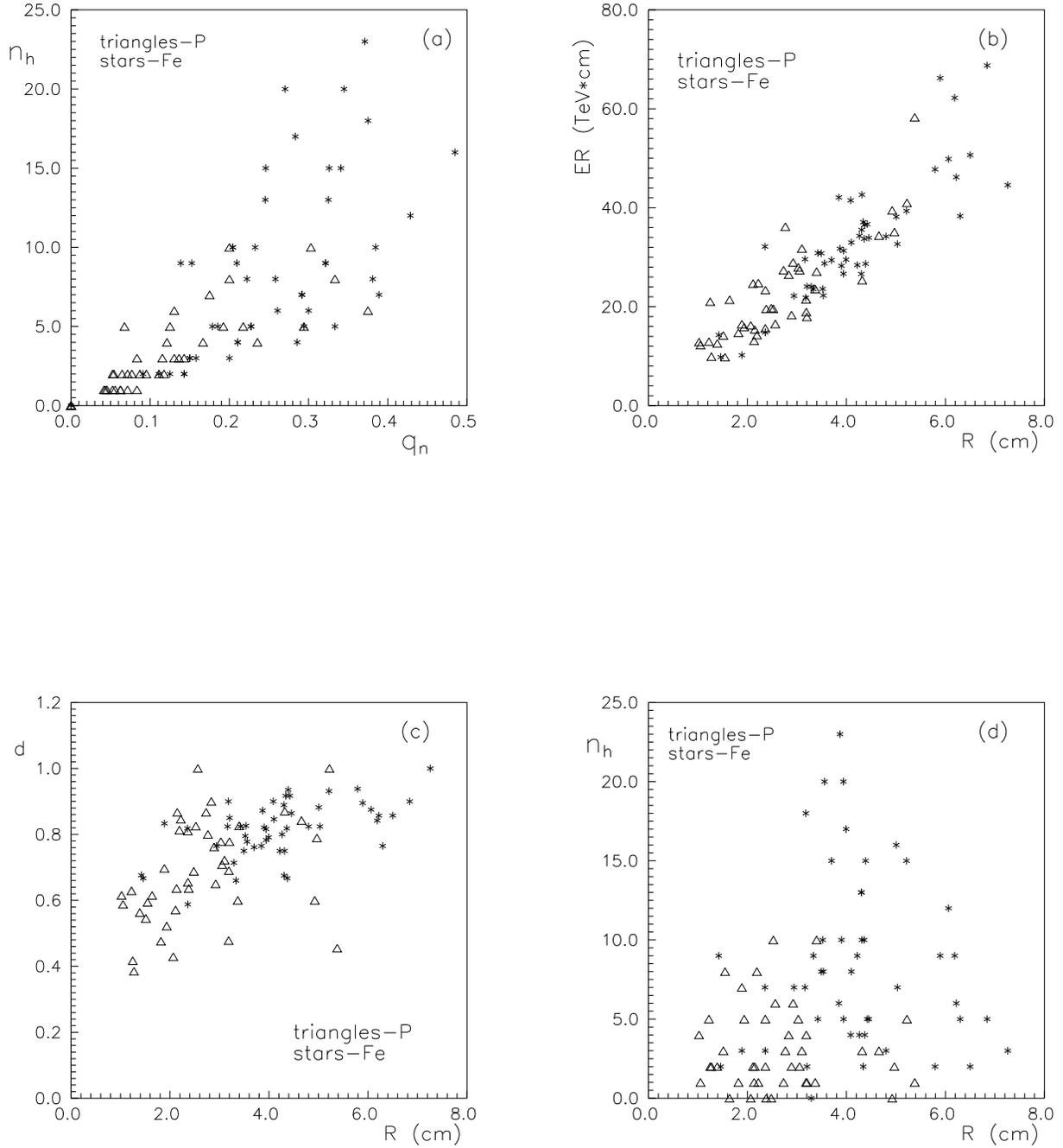}
\caption{Correlation: a)between $n_{h}$ and
$q_{h}$,~k=0.60$\pm$0.02;~b)between $R_{\gamma}$ and
$ER_{\gamma}$,~k=0.88$\pm$0.01;~~ c) between d and
$R_{\gamma}$,~k=0.34$\pm$0.03;~~ d) between $n_{h}$ and
$R_{\gamma}$,}~k=0.00$\pm$0.04.
\end{figure}

\begin{figure}
\includegraphics{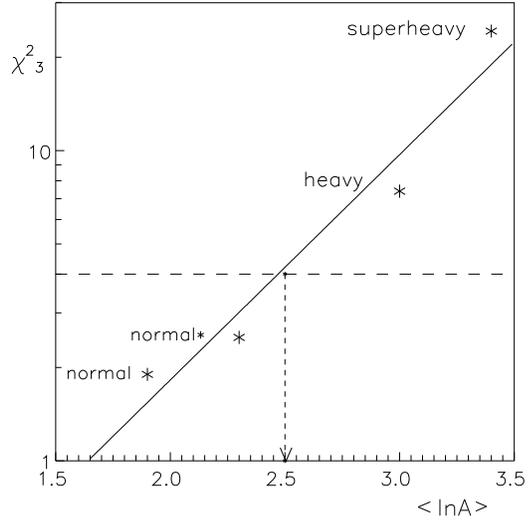}
\caption{Dependence of $\chi^2_3$ on $\langle\ln{A}\rangle$.}
\end{figure}
\begin{figure}
\includegraphics{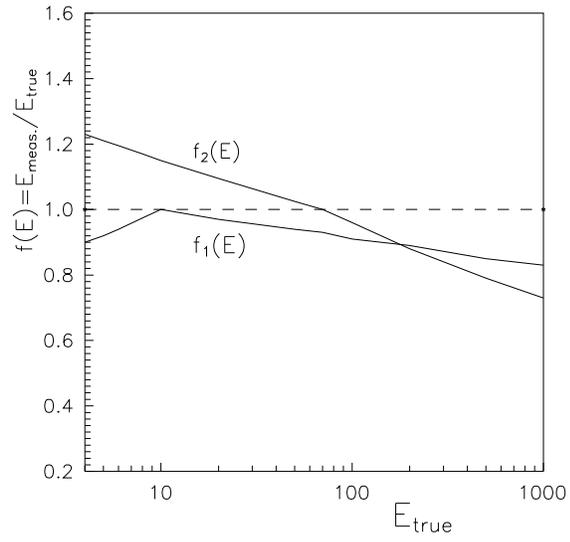}
\caption{Two types of distortion functions.}
\end{figure}

\begin{figure}
\includegraphics{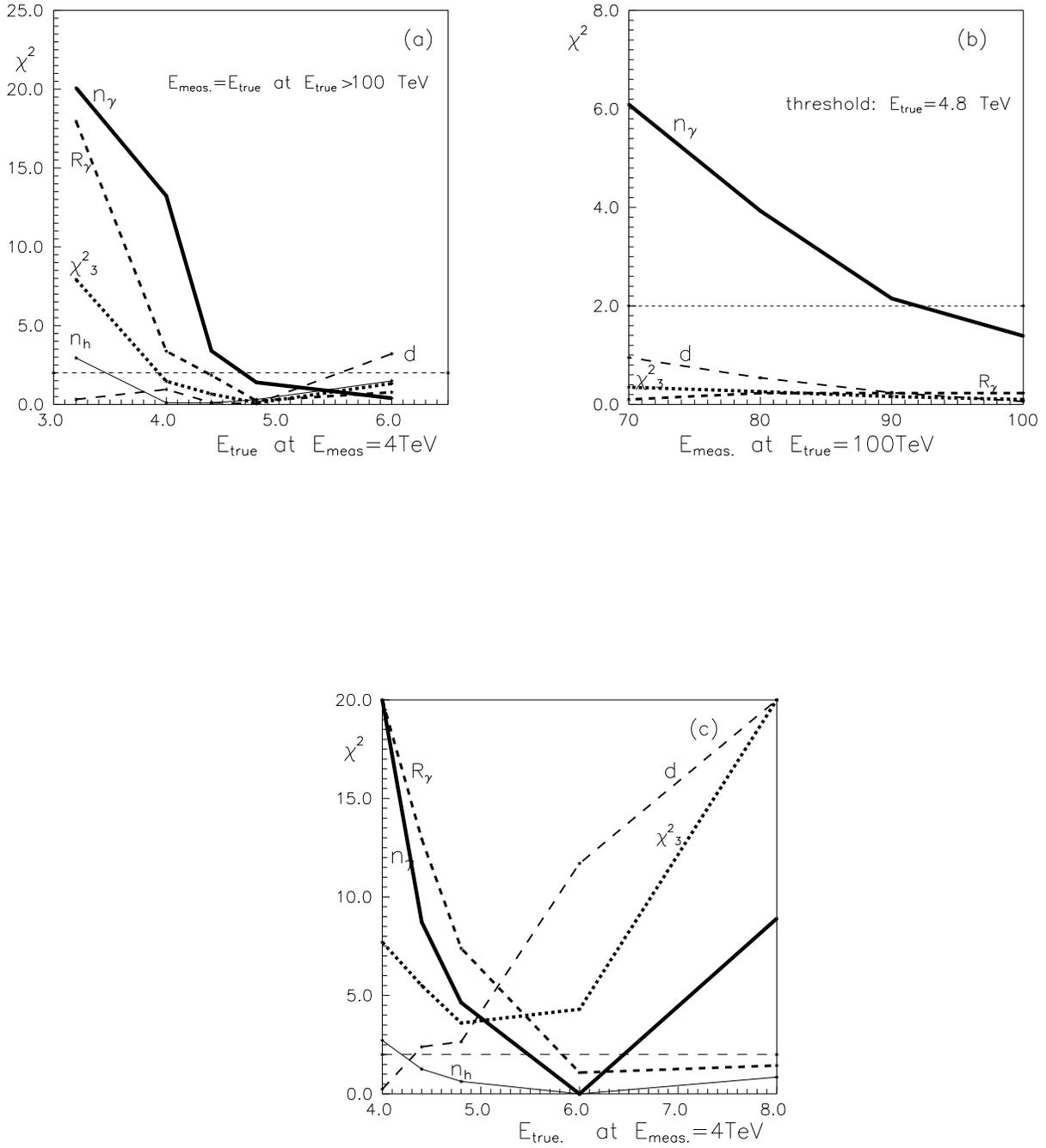}
\caption{Dependencies of $\chi^2_P$ and $\chi^2_3$ on: a) degree
of energy underestimation near measured threshold 4~ TeV at normal
chemical composition, b) degree of energy overestimation near true
energy 100~TeV at normal chemical composition , c) the same as a)
for heavy CC.}
\end{figure}

\begin{figure}
\includegraphics{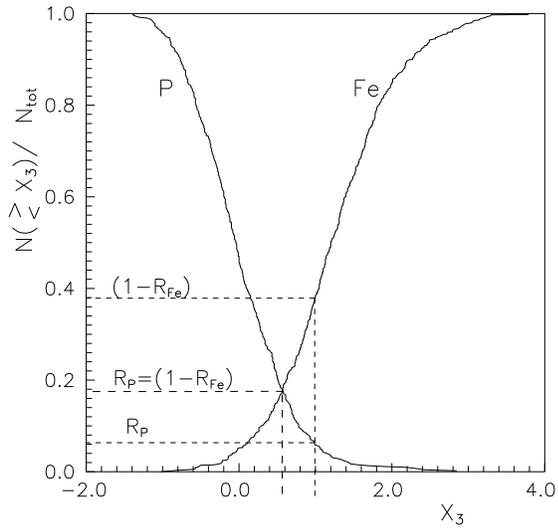}
\caption{Integral distributions of $X_{3}$  for families,
generated by protons and iron. Limited values of $X_{3}$
correspondent to conditions $R_{P} = R_{Fe}$  and $R_{P}$ = 5$\%$,
are shown by dashed lines. }
\end{figure}

\end{document}